# Kinematics and Velocity Ellipsoid of the Solar Neighborhood White Dwarfs

W. H. Elsanhoury[1,2], M. I. Nouh[1,3] and H. I. Abdel-Rahman[1,4]

[1]Astronomy Dept., National Research Institute of Astronomy and Geophysics (NRIAG) 11421, Helwan, Cairo, EGYPT, Email: welsanhoury@gmail.com

[2]Department of Physics, Faculty of Science and Arts, Northern Border University, Rafha Branch, Saudi Arabia

[3]Department of Physics, College of Science, Northern Border University, 1321 Arar, Saudi Arabia

[4]Department of Mathematics, Faculty of Science, Shaqraa University, Shaqraa, Saudi Arabia

**Abstract:** To determine the velocity ellipsoid of the solar neighborhood white dwarfs, we use the space velocity components of stars. Two samples of white dwarfs are used, 20 pc and 25 pc samples. Beside the two main samples, the solar velocity and velocity dispersions are calculated for the four subsamples, namely DA, non - DA, hot and cool white dwarfs. Comparison between the results of 20 pc sample and those of 25 pc sample gives good agreement, while the comparison between the other subsamples gives bad agreement. Dependence of the velocity dispersions and solar velocity on the chemical composition and effective temperatures are discussed.

## 1. Introduction

Majority of stars will eventually end their lives as white dwarfs. These faint stellar remnants can be used in many different investigations in astrophysics. White dwarf cooling process has been used to date the globular star cluster M4 (Hansen et al., 2004; Hansen et al., 2002) and independently determines the age of the Galactic halo. Also white dwarfs are used to determine the mass function of the cluster above the main-sequence turn-off (Richer et al., 2004 and Richer et al., 2002). Since all stars with a mass above $0.8\, M_\odot$ have evolved off the main-sequence in a 12 Gyr population, the white dwarfs represent our only link to the distribution of stars (i.e., the initial mass function) of intermediate and massive stars in these systems. White dwarfs are also astrophysically important when considering the chemical evolution of the Galaxy.

The velocity distribute on of stars in the solar neighborhood has been characterized as an ellipsoid the centroid, size, and orientation of which vary systematically with the ages (and hence colors) of the stars under investigation, Hogg et al. (2005) and Dehnen and Binney (1998).

It is well known since a long time (Ogorpdnikov, 1965) that, in the neighborhood of the Sun, the characteristic feature of stellar motion is the fact that the peculiar velocities have an axis of greatest mobility and this characteristic is represented most conveniently on the basis of ellipsoidal law of velocity distribution.

In the present paper, we are going to determine the velocity ellipsoid of solar neighborhood white dwarfs. We investigate the dependence of the velocity parameters on the number of stars, spectral type and effective temperatures. The structure of the paper is as follows: section 1 deals with the method of computation and data used. Section 3 is devoted to the results and discussions. The conclusion is outlined in section 4.

## 2. Data and Method of Computation

### 2.1 Data

The data used in the present computations are due to Sion et al. (2009) and Sion et al. (2014) for white dwarf within 20 and 25 pc of the Sun. The total number of the 20 pc sample contains 126 candidates of different spectral types of white dwarfs.

The 25 pc sample contains 141 of spectral type DA and 68 of non-DA. The effective temperature ranges from 2600 K° to 30510 K°. The vector components of the space motions *U*, *V* and *W* are computed and tabulated.

The atmospheric parameter in the two samples were determined by different methods; i.e. photometric, spectroscopic and parallax observations.

In Table (1) we listed the 25 pc white dwarfs list (209 candidate). The columns are designed as follows: the WD-number in column 1, the spectral type in column 2, the effective temperature in column 3 and the space motions $U$, $V$ and $W$ in column 4, and 6 respectively.

**Table 1:** Data of 209 WD stars (25 pc) devoted by Sion et al. (2014)

| No. | WD No. | Sp. Type | $T_{eff}$ | U | V | W | No. | WD No. | Sp. Type | $T_{eff}$ | U | V | W |
|---|---|---|---|---|---|---|---|---|---|---|---|---|---|
| 1 | 0000-345 | DCP8.1 | 6643 | -11.9 | -43.7 | 3.2 | 106 | 1121+216 | DA6.7 | 7434 | 57.8 | -25.3 | -21.5 |
| 2 | 0008+424 | DA6.8 | 7380 | -10 | -1.9 | -17.4 | 107 | 1124+595 | DA4.8 | 10747 | -12.2 | 1.3 | 6 |
| 3 | 0009+501 | DAH7.6 | 6502 | -28.1 | 8.6 | -24.1 | 108 | 1134+300 | DA2.4 | 22469 | 9.2 | -4.6 | -2.9 |
| 4 | 0011-134 | DAH8.4 | 5992 | -75.5 | -31.1 | -9.8 | 109 | 1142-645 | DQ6.4 | 7966 | -52.1 | 25.3 | 6.6 |
| 5 | 0011-721 | DA7.8 | 6325 | 10.3 | -23.2 | 12.2 | 110 | 1149-272 | DQ8.1 | 6200 | 24.9 | -8.5 | -0.8 |
| 6 | 0029-031 | DA11.3 | 4470 | 68.3 | -23 | 2.6 | 111 | 1202-232 | DAZ5.8 | 8767 | 3 | 6.1 | 8.6 |
| 7 | 0038+555 | DQ4.6 | 10900 | 39.4 | -2 | -11.9 | 112 | 1208+576 | DAZ8.6 | 6200 | -45.5 | -10.6 | 23 |
| 8 | 0038-226 | DQpec9.3 | 5529 | -24.9 | -4.6 | -1.2 | 113 | 1214+032 | DA8.0 | 6272 | 64 | -11.2 | 2.3 |
| 9 | 0046+051 | DZ7.4 | 6215 | -2.8 | -53.6 | -30.3 | 114 | 1223-659 | DA6.6 | 7594 | 0.7 | 0.1 | -8.4 |
| 10 | 0108+048 | DA6.4 | 8530 | 38.2 | -0.1 | 11.8 | 115 | 1236-495 | DA4.3 | 11599 | 24 | -17 | -8.6 |
| 11 | 0108+277 | DA9.6 | 6428 | -11.9 | 0.3 | -9.9 | 116 | 1241-798 | DC/DQ | 9556 | 38.1 | -36.9 | 33.3 |
| 12 | 0115+159 | DQ5.6 | 9119 | -21.4 | -27.2 | -32 | 117 | 1257+037 | DA9.0 | 5616 | -6.2 | -69.4 | -25.6 |
| 13 | 0121-429 | DAH7.9 | 6299 | -0.7 | -46.8 | 11.1 | 118 | 1309+853 | DAP9 | 5440 | -16.3 | -1.3 | 15.4 |
| 14 | 0123-460 | DA8.5 | 5898 | 24.3 | -91 | 24.9 | 119 | 1310+583 | DA4.8 | 10544 | -16.3 | 6.7 | 5.7 |
| 15 | 0134+883 | DA2.8 | 18311 | -11.1 | 5.1 | 6.1 | 120 | 1310-472 | DC11.9 | 4158 | -133.6 | 81.4 | -51.5 |
| 16 | 0135-052 | DA6.9 | 7118 | 16.8 | -37.2 | -1.6 | 121 | 1315-781 | DC8.8 | 5619 | -30.2 | 34 | -41.7 |
| 17 | 0141-675 | DA7.8 | 6248 | -21.8 | -22.8 | 21.5 | 122 | 1315-781 | DC8.8 | 5619 | -23.5 | 26.4 | -32.3 |
| 18 | 0148+467 | DA3.8 | 14005 | 2.6 | 1.8 | 8.3 | 123 | 1327-083 | DA3.6 | 14571 | 49.2 | -76.6 | -7.4 |
| 19 | 0148+641 | DA5.6 | 9016 | 12.3 | -11.2 | -6.8 | 124 | 1334+039 | DA11 | 4971 | 87.5 | -122.2 | 8.5 |
| 20 | 0208+396 | DAZ6.9 | 7264 | 43 | -62.9 | -6.7 | 125 | 1344+106 | DAH7.1 | 7059 | 54.6 | -62.3 | 14.1 |
| 21 | 0213+427 | DA9.0 | 5507 | 37.8 | -67.3 | -19.8 | 126 | 1337+705 | DAZ2.5 | 20464 | 43.7 | -15.5 | 29.9 |
| 22 | 0227+050 | DA2.7 | 18779 | -0.1 | -8.9 | 7.1 | 127 | 1339-340 | DA9.5 | 5361 | 55 | 108 | 177.4 |
| 23 | 0230-144 | DA9.2 | 5477 | -20.7 | -43.3 | -13.3 | 128 | 1344+572 | DA3.8 | 13389 | 27.3 | 26.8 | 37.5 |
| 24 | 0231-054 | DA3.7 | 13550 | 68.5 | -6.7 | -56.4 | 129 | 1345+238 | DA11 | 4581 | 67 | -47.4 | 20.1 |
| 25 | 0233-242 | DC9.3 | 5312 | -24.2 | -34.1 | -9.8 | 130 | 1350-090 | DAP5 | 9518 | -21.7 | -23 | -23.2 |
| 26 | 0236+259 | DA9.2 | 5500 | 29.1 | -21.1 | -8.9 | 131 | 1401+457 | DC19 | 2600 | 16.9 | -27.6 | 5.5 |
| 27 | 0243-026 | DAZ7.4 | 6839 | 8.9 | -51.6 | -34.5 | 132 | 1425-811 | DAV4.2 | 12098 | -7.9 | -26.2 | -44 |
| 28 | 0245+541 | DAZ9.5 | 5319 | -16.4 | 10.9 | -24.2 | 133 | 1436-781 | DA8.1 | 6270 | 29.7 | -32.5 | 23.3 |
| 29 | 0255-705 | DAZ4.7 | 10560 | 22.2 | -82 | -0.6 | 134 | 1444-174 | DC10.2 | 4982 | 33 | -61.4 | 17.6 |
| 30 | 0310-688 | DA3.3 | 16865 | 0.1 | -4.2 | 2.8 | 135 | 1532+129 | DZ6.7 | 7500 | -1.8 | -26.2 | 3.9 |
| 31 | 0311-649 | DA4.0 | 11945 | 7.9 | -12.8 | 9.5 | 136 | 1538+333 | DA5.6 | 8940 | 12 | -5.2 | 11.8 |
| 32 | 0322-019 | DAZ9.9 | 5195 | -21.5 | -66.8 | -20.7 | 137 | 1544-377 | DA4.8 | 10610 | 5.5 | -22.4 | 7.6 |
| 33 | 0326-273 | DA5.4 | 8483 | 46.7 | -29.7 | 47.2 | 138 | 1609+135 | DA5.4 | 9041 | -25.2 | -35.3 | -17.4 |
| 34 | 0341+182 | DQ7.7 | 6568 | -16 | -94.9 | -23.8 | 139 | 1620-391 | DA2.1 | 25985 | -1.3 | 2.8 | -3.2 |
| 35 | 0344+014 | DC9.9 | 5170 | -7.9 | -43.8 | -4.9 | 140 | 1625+093 | DA7.3 | 7038 | -23.9 | -46.7 | -12.3 |
| 36 | 0357+081 | DA9.2 | 5478 | -25.3 | -4.3 | -36.8 | 141 | 1626+368 | DZA6.0 | 8507 | 35.1 | 15.5 | 35.6 |
| 37 | 0413-077 | DA3.1 | 17100 | -44.2 | -34.8 | -73 | 142 | 1632+177 | DAZ5.0 | 10225 | -2.7 | 2.3 | -5.3 |
| 38 | 0416-594 | DA3.3 | 14000 | -4.9 | -5.7 | -1.6 | 143 | 1633+433 | DAZ7.7 | 6608 | -12.7 | -4.2 | -13.9 |
| 39 | 0419-487 | DA8 | 6300 | -1.2 | -91.1 | -57.4 | 144 | 1633+572 | DQ8.2 | 5958 | 45.4 | -3 | 54.5 |
| 40 | 0423+044 | DA | 5140 | -6.9 | -80.5 | 17.2 | 145 | 1639+537 | DAH6.7 | 7510 | -3.6 | -19.1 | 8.9 |
| 41 | 0423+120 | DA8.2 | 6167 | 5.3 | 18.1 | 3.8 | 146 | 1647+591 | DAV4.1 | 12738 | -6 | -3 | -5.4 |
| 42 | 0426+588 | DC7.1 | 7178 | 0.9 | -35.4 | -3.7 | 147 | 1655+215 | DAB5.4 | 9179 | -38 | -39 | -17.7 |
| 43 | 0431-279 | DC9.5 | 5330 | 11.5 | -28.8 | 33.3 | 148 | 1655+215 | DAB5.4 | 9179 | -25.9 | -35.9 | -18.4 |
| 44 | 0431-360 | DA10.0 | 5153 | 10.8 | -19.6 | 27 | 149 | 1658+440 | DAP1.7 | 30510 | -6.6 | 33 | 28.2 |
| 45 | 0433+270 | DA9.0 | 5629 | -0.1 | -19.8 | 7 | 150 | 1705+030 | DZ7.7 | 6584 | -16.6 | -23.1 | -13.4 |
| 46 | 0435-088 | DQ8.0 | 6367 | -25.1 | -63.2 | -21.6 | 151 | 1748+708 | DQ9.0 | 5570 | 5.8 | -10.8 | 34.2 |
| 47 | 0457-004 | DA4.7 | 10800 | -6.1 | -23.8 | 2.5 | 152 | 1756+143 | DA9.0 | 5466 | -36.3 | -86.3 | 49.1 |
| 48 | 0503-174 | DAH9.5 | 5300 | 32.2 | 42.5 | 41 | 153 | 1756+827 | DA6.9 | 7214 | 8.8 | -30.3 | 106.4 |
| 49 | 0511+079 | DA7.7 | 6590 | -14.5 | -9.2 | -30.6 | 154 | 1814+134 | DA9.5 | 5251 | -47.7 | -72 | -8.6 |
| 50 | 0532+414 | DA6.8 | 7739 | 1.2 | 12.4 | -13.4 | 155 | 1817-598 | DA5.8 | 4960 | -15.1 | -22.9 | 2.3 |
| 51 | 0548-001 | DQP8.3 | 6070 | 4.6 | 6.3 | 10.5 | 156 | 1820+609 | DA10.5 | 4919 | -9.8 | -9 | -19.7 |
| 52 | 0552-041 | DZ10.0 | 5182 | -25.3 | -65.7 | -20.2 | 157 | 1829+547 | DQP8.0 | 6345 | 2.8 | -0.7 | 24.3 |
| 53 | 0553+053 | DAP8.7 | 5785 | -12.2 | -19.8 | -31.2 | 158 | 1840+042 | DA5.8 | 9090 | 32 | -23.6 | 19.9 |
| 54 | 0615-591 | DB3.2 | 16714 | -11.3 | -0.1 | 7.1 | 159 | 1900+705 | DAP4.2 | 11835 | 6.7 | 5.5 | 6 |
| 55 | 0618+067 | DA8.1 | 5940 | -9 | -26.9 | 49.9 | 160 | 1917+386 | DC7.9 | 6459 | -5.2 | -5.7 | -8.6 |
| 56 | 0620-402 | DZ6 | 5919 | -13.9 | -25 | -7.2 | 161 | 1917-077 | DBQZ4.9 | 10396 | -5.4 | -6.4 | -0.5 |

| | | | | | | | | | | | | |
|---|---|---|---|---|---|---|---|---|---|---|---|---|
| 57 | 0628-020 | DA | 6912 | 61 | -34.5 | -27.2 | 162 | 1919+145 | DA3.3 | 15280 | -4.2 | -5 | -0.7 |
| 58 | 0628-020 | DA | 6912 | -5.2 | -7.7 | -17.7 | 163 | 1935+276 | DA4.2 | 12130 | 16 | 10.2 | -31.7 |
| 59 | 0642-166 | DA2 | 25967 | -0.6 | -10.2 | -14.5 | 164 | 1953-011 | DC6.4 | 7920 | -32.3 | -31.9 | 3.7 |
| 60 | 0644+025 | DA6.8 | 22288 | 9.4 | 15.1 | -30.9 | 165 | 2002-110 | DA10.5 | 4800 | 40.4 | 9.8 | -76.6 |
| 61 | 0644+375 | DA2.4 | 22288 | -19.7 | -39.6 | -32.7 | 166 | 2007-303 | DA3.5 | 14454 | -22 | -18.3 | 18 |
| 62 | 0651-398A | DA7.0 | 7222 | 5.2 | 28.7 | 8.9 | 167 | 2008-600 | DC9.9 | 5080 | -17.7 | -62.3 | -3.3 |
| 63 | 0655-390 | DA7.9 | 6311 | 2.7 | 2.3 | -28.2 | 168 | 2008-799 | DA8.5 | 5800 | 18 | -6.5 | -20.7 |
| 64 | 0657+320 | DA10.1 | 4888 | -31.7 | -52.9 | 13.2 | 169 | 2011+065 | DQ7 | 6400 | -64.2 | -24.8 | -17 |
| 65 | 0659-063 | DA7.7 | 6627 | -19.7 | -37.5 | -29.5 | 170 | 2032+248 | DA2.4 | 19983 | -36.9 | -24.9 | -2.8 |
| 66 | 0706+377 | DQ7.6 | 6590 | -7.2 | -13 | -31.6 | 171 | 2039-202 | DA2.5 | 19207 | 15.9 | -3.9 | -30.9 |
| 67 | 0708-670 | DC9.9 | 5097 | 4.7 | 4.9 | -19.9 | 172 | 2039-682 | DA3.1 | 15855 | 0.7 | -17.8 | -9.9 |
| 68 | 0727+482.1 | DA10.0 | 4934 | -14.6 | -40.5 | -14.7 | 173 | 2040-392 | DA4.5 | 10830 | -13.6 | -25.2 | -0.3 |
| 69 | 0727+482.2 | DA10.1 | 4926 | -14.9 | -41.1 | -14.9 | 174 | 2047+372 | DA3.6 | 14070 | 13.4 | 5.5 | -1.4 |
| 70 | 0728+642 | DAP11.1 | 5135 | -4.5 | -9.2 | 3.9 | 175 | 2048+263 | DA9.7 | 5200 | -79.5 | 104.1 | -13.4 |
| 71 | 0736+053 | DQZ6.5 | 7871 | 0.6 | -12 | -18.4 | 176 | 2048+263 | DA9.7 | 5200 | -37.6 | -16.2 | 11.8 |
| 72 | 0738-172 | DZA6.6 | 7650 | -30.5 | -29.2 | 30.2 | 177 | 2048-250 | DA6.6 | 7630 | 5.3 | -13.6 | -20 |
| 73 | 0743-336 | DC10.6 | 4462 | 49.4 | 67.6 | 61.4 | 178 | 2054-050 | DC10.9 | 4620 | 31.1 | -11.4 | -54.1 |
| 74 | 0747+073.1 | DC10.4 | 4366 | -76.7 | -126.3 | -49 | 179 | 2105-820 | DA4.7 | 10620 | 12.5 | -17.4 | 3.6 |
| 75 | 0747+073.2 | DC12.0 | 4782 | -76.7 | -126.3 | -49 | 180 | 2115-560 | DA6 | 9736 | 17.5 | -18.6 | -31.4 |
| 76 | 0749+426 | DC11.7 | 4585 | -17.8 | -29.6 | 5.6 | 181 | 2117+539 | DA3.6 | 13990 | -0.8 | 2.1 | 18.3 |
| 77 | 0751-252 | DA9.8 | 5085 | 19.1 | 15.5 | -13.8 | 182 | 2118-388 | DC9.6 | 5244 | 9.3 | -6.3 | -11.6 |
| 78 | 0752-676 | DA8.8 | 5735 | -29.3 | -15.4 | 13.6 | 183 | 2126+734 | DA3.8 | 16104 | 3.2 | 11.8 | -23.5 |
| 79 | 0753+417 | DA7.3 | 6880 | -13.8 | -28.6 | -3.5 | 184 | 2133-135 | DA5.0 | 9736 | 11.7 | -12.8 | -20.3 |
| 80 | 0805+356 | DA7.3 | 6900 | 0.5 | -4.9 | -6.6 | 185 | 2138-332 | DZ7 | 7240 | -15 | -8.3 | 8.4 |
| 81 | 0806-661 | DQ4.9 | 10205 | -17.2 | -6.5 | 7.3 | 186 | 2140+207 | DQ6.1 | 8200 | -28 | -18.9 | -17.1 |
| 82 | 0810+489 | DC6.9 | 7300 | -14.5 | -22.8 | 8.6 | 187 | 2149+021 | DA2.8 | 17353 | -28.4 | -3.2 | -37.7 |
| 83 | 0810+489 | DC6.9 | 7300 | -9 | -15.1 | 5 | 188 | 2151-015 | DA6 | 8400 | -59 | 52.4 | -85.8 |
| 84 | 0816-310 | DZ7.6 | 6463 | -45 | -52.7 | -37 | 189 | 2154-512 | DQ8.3 | 6100 | -12.4 | -22.4 | 9.9 |
| 85 | 0821-669 | DA9.8 | 5088 | 16.8 | 4.5 | 4.4 | 190 | 2159-754 | DA5.6 | 9040 | -27.6 | 7.6 | 18.5 |
| 86 | 0827+328 | DA6.9 | 7490 | 4.1 | -45.4 | -6.6 | 191 | 2211-392 | DA8.1 | 6920 | 46.6 | -43.1 | -61.1 |
| 87 | 0839-327 | DA5.5 | 9081 | 38.2 | 26.6 | 4.6 | 192 | 2211-392 | DA8.1 | 6920 | 57.8 | -43.7 | -44.5 |
| 88 | 0840-136 | DZ10.3 | 4874 | 14 | 0 | -21 | 193 | 2211-392 | DA8.1 | 6920 | 46.6 | -43.1 | -61.1 |
| 89 | 0843+358 | DZ6 | 9041 | 7.6 | -5.5 | -13.6 | 194 | 2211-392 | DA8.1 | 6920 | 57.8 | -43.7 | -44.5 |
| 90 | 0856+331 | DQ5.1 | 9920 | 21.4 | 0.9 | -24.3 | 195 | 2215+386 | DC10.6 | 4700 | 48.7 | -7.1 | -20.6 |
| 91 | 0912+536 | DCP7 | 7235 | 21 | -42.5 | -24.3 | 196 | 2226-754 | DC11.9 | 4230 | -0.5 | -88.5 | 71.5 |
| 92 | 0946+534 | DQ6.2 | 8100 | 20.2 | -7.3 | -16.1 | 197 | 2226-755 | DC12.1 | 4177 | -0.5 | -88.5 | 71.5 |
| 93 | 0955+247 | DA5.8 | 8621 | 19.1 | -40.3 | -11.8 | 198 | 2246+223 | DA4.7 | 10647 | 42.4 | -10.4 | -16.9 |
| 94 | 0955+247 | DA5.8 | 8621 | 6.1 | -28.1 | -17.7 | 199 | 2248+293 | DA9 | 5580 | 111.2 | -29.8 | -43 |
| 95 | 0959+149 | DC7 | 7200 | -39.4 | 56.1 | 127.2 | 200 | 2251-070 | DZ12.6 | 4000 | 68 | -48.2 | -50 |
| 96 | 1009-184 | DZ8.5 | 6036 | 35.1 | -6 | -26.8 | 201 | 2253+054 | DA9 | 5600 | 20.3 | -33.8 | -34.2 |
| 97 | 1009-184 | DZ8.5 | 6036 | 36.1 | -8.8 | -26.4 | 202 | 2311-068 | DQ6.8 | 7440 | -45.3 | -1.7 | 7.4 |
| 98 | 1012+083.1 | DA7.5 | 6750 | 38.4 | 10.1 | -12.2 | 203 | 2322+137 | DA10.7 | 4700 | 3.2 | -0.5 | -0.3 |
| 99 | 1019+637 | DA7.2 | 6742 | -14.7 | 17.9 | 3.5 | 204 | 2326+049 | DAV4.3 | 12206 | -31.6 | -2.1 | -1.3 |
| 100 | 1033+714 | DC10.3 | 4727 | 138.8 | -73.2 | -58 | 205 | 2336-079 | DA4.6 | 10938 | -5.8 | -13.4 | -5.9 |
| 101 | 1036-204 | DQP10.2 | 4694 | 36 | 17.9 | 20.7 | 206 | 2341+322 | DA4.0 | 13128 | -18.8 | 5.6 | -0.4 |
| 102 | 1043-188 | DQ8.1 | 5780 | 110.7 | -71.1 | 107.8 | 207 | 2347+292 | DA9 | 5810 | -18.3 | 2.7 | -57.2 |
| 103 | 1055-072 | DA6.8 | 7491 | 42.3 | -10.4 | -16.3 | 208 | 2351-335 | DA5.7 | 8850 | -62 | -22.5 | -38.2 |
| 104 | 1105-048 | DA3.5 | 15141 | -17.3 | -37.8 | -29.8 | 209 | 2359-434 | DA5.9 | 8648 | 13.9 | -16.4 | -27.6 |
| 105 | 1116-470 | DC8.6 | 5801 | 24.3 | -9.2 | -7 | | | | | | | |

## 2.2 Model

To compute the velocity ellipsoid and velocity ellipsoid parameters of the solar neighborhood white dwarfs, we follow the computational algorithm of Elsanhoury et al. (2013). Brief explanation of the algorithm will be given here.

The coordinates of the $i^{th.}$ star with respect to axes parallel to the original axes, but shifted into the center of the distribution, i.e. into the point $\overline{U}, \overline{V}$ and $\overline{W}$, will be $(U_i - \overline{U}); (V_i - \overline{V}); (W_i - \overline{W})$. Where the components $U$, $V$ and $W$ are the components of the space velocities and $\overline{U}, \overline{V}$ and $\overline{W}$ are the mean velocities defined as:

$$\overline{U}=\frac{1}{N}\sum_{i=1}^{N}U_i;\ \overline{V}=\frac{1}{N}\sum_{i=1}^{N}V_i;\ \overline{W}=\frac{1}{N}\sum_{i=1}^{N}W_i \qquad (1)$$

$N$ being the total number of the stars.

If we have an arbitrary axis $\xi$ (say) and its zero point coincide with the center of the distribution and let $l, m$ and $n$ be the direction cosines of the axis with respect to the shifted one, then the coordinates $Q_i$ of the point $i$, with respect to the $\xi$ axis is given by:

$$Q_i = l\left(U_i - \overline{U}\right) + m\left(V_i - \overline{V}\right) + n\left(W_i - \overline{W}\right). \qquad (2)$$

Let us adopt, as the measured of the scatter components $Q_i$ a generalization of the mean square deviation, defined by

$$\sigma^2 = \frac{1}{N}\sum_{i=1}^{N}Q_i^2 \qquad (3)$$

From Equations (1), (2) and (3) we deduce after some calculations that

$$\sigma^2 = \underline{x}^T B \underline{x} \qquad (4)$$

where $\underline{x}$ is the $(3 \times 1)$ direction cosines vector and $B$ is $(3 \times 3)$ symmetric matrix $\mu_{ij}$ and

$$\left.\begin{array}{l} \mu_{11} = \frac{1}{N}\sum_{i=1}^{N}U_i^2 - \left(\overline{U}\right)^2; \quad \mu_{12} = \frac{1}{N}\sum_{i=1}^{N}U_i V_i - \overline{U}\,\overline{V}; \\ \mu_{13} = \frac{1}{N}\sum_{i=1}^{N}U_i W_i - \overline{U}\,\overline{W}; \quad \mu_{22} = \frac{1}{N}\sum_{i=1}^{N}V_i^2 - \left(\overline{V}\right)^2; \\ \mu_{23} = \frac{1}{N}\sum_{i=1}^{N}V_i W_i - \overline{V}\,\overline{W}; \quad \mu_{33} = \frac{1}{N}\sum_{i=1}^{N}W_i^2 - \left(\overline{W}\right)^2. \end{array}\right\} \qquad (5)$$

$\mu_{ij}$ are the matrix elements. The necessary conditions for an extremum are now

$$(B - \lambda I)\underline{x} = 0 \qquad (6)$$

These are three homogenous equations in three unknowns have a nontrivial solution if and only if

$$D(\lambda) = |B - \lambda I| = 0, \tag{7}$$

where $\lambda$ is the eigenvalue, and $\underline{x}$ and $B$ are given as:

$$\underline{x} = \begin{bmatrix} l \\ m \\ n \end{bmatrix} \text{ and } B = \begin{vmatrix} \mu_{11} & \mu_{12} & \mu_{13} \\ \mu_{12} & \mu_{22} & \mu_{23} \\ \mu_{13} & \mu_{23} & \mu_{33} \end{vmatrix}$$

Equation (7) is characteristic equation for the matrix $B$. The required roots (i.e. eigenvalues) are

$$\left. \begin{aligned} \lambda_1 &= 2\rho^{\frac{1}{3}} \cos\frac{\phi}{3} - \frac{k_1}{3}; \\ \lambda_2 &= -\rho^{\frac{1}{3}} \left\{ \cos\frac{\phi}{3} + \sqrt{3}\sin\frac{\phi}{3} \right\} - \frac{k_1}{3}; \\ \lambda_3 &= -\rho^{\frac{1}{3}} \left\{ \cos\frac{\phi}{3} - \sqrt{3}\sin\frac{\phi}{3} \right\} - \frac{k_1}{3}. \end{aligned} \right\} \tag{8}$$

where

$$\left. \begin{aligned} k_1 &= -(\mu_{11} + \mu_{22} + \mu_{33}), \\ k_2 &= \mu_{11}\mu_{22} + \mu_{11}\mu_{33} + \mu_{22}\mu_{33} - (\mu_{12}^2 + \mu_{13}^2 + \mu_{23}^2), \\ k_3 &= \mu_{12}^2\mu_{33} + \mu_{13}^2\mu_{22} + \mu_{23}^2\mu_{11} - \mu_{11}\mu_{22}\mu_{33} - 2\mu_{12}\mu_{13}\mu_{23}. \end{aligned} \right\} \tag{9}$$

$$q = \frac{1}{3}k_2 - \frac{1}{9}k_1^2 \quad ; \quad r = \frac{1}{6}(k_1 k_2 - 3k_3) - \frac{1}{27}k_1^3 \tag{10}$$

$$\rho = \sqrt{-q^3} \tag{11}$$

$$x = \rho^2 - r^2 \tag{12}$$

and

$$\phi = \tan^{-1}\left(\frac{\sqrt{x}}{r}\right) \tag{13}$$

Depending on the matrix that control the eigenvalue problem [Equation (6)] for the velocity ellipsoid, we establish analytical expressions of some parameters for the correlations studies in terms of the matrix elements $\mu_{ij}$ of the eigenvalue problem for the velocity ellipsoid (i.e. Velocity Ellipsoid Parameters VEPs).

- **The $\sigma_i$; $i = 1, 2, 3$ parameters**

The $\sigma_i$; $i = 1, 2, 3$ parameters are defined as

$$\sigma_i = \sqrt{\lambda_i} \tag{14}$$

- **The $l_i$, $m_i$ and $n_i$ parameters**

The $l_i$, $m_i$ and $n_i$ are the direction cosines for eigenvalue problem. Then we have the following expressions for $l_i$, $m_i$ and $n_i$ as

$$l_i = \left[\mu_{22}\mu_{33} - \sigma_i^2\left(\mu_{22} + \mu_{33} - \sigma_i^2\right) - \mu_{23}^2\right]/D_i \quad ; i = 1, 2, 3 \tag{15}$$

$$m_i = \left[\mu_{23}\mu_{13} - \mu_{12}\mu_{33} + \sigma_i^2\mu_{12}\right]/D_i \quad ; i = 1, 2, 3 \tag{16}$$

$$n_i = \left[\mu_{12}\mu_{23} - \mu_{13}\mu_{22} + \sigma_i^2\mu_{13}\right]/D_i \quad ; i = 1, 2, 3 \tag{17}$$

where

$$\begin{aligned}
D_i^2 &= \left(\mu_{22}\mu_{33} - \mu_{23}^2\right)^2 + \left(\mu_{23}\mu_{13} - \mu_{12}\mu_{33}\right)^2 + \left(\mu_{12}\mu_{23} - \mu_{13}\mu_{22}\right)^2 \\
&+ 2\left[\left(\mu_{22} + \mu_{33}\right)\left(\mu_{23}^2 - \mu_{22}\mu_{33}\right) + \mu_{12}\left(\mu_{23}\mu_{13} - \mu_{12}\mu_{33}\right) + \mu_{13}\left(\mu_{12}\mu_{23} - \mu_{13}\mu_{22}\right)\right]\sigma_i^2 \\
&+ \left(\mu_{33}^2 + 4\mu_{22}\mu_{33} + \mu_{22}^2 - 2\mu_{23}^2 + \mu_{12}^2 + \mu_{13}^2\right)\sigma_i^4 - 2\left(\mu_{22} + \mu_{33}\right)\sigma_i^6 + \sigma_i^8.
\end{aligned}$$

## 3. Results

Based on the model described in the previous section, Mathematica routine has been developed to compute the kinematics and velocity ellipsoid parameters. Figure 1 shows the distribution of the space velocities of 209 white dwarfs (25 pc). Beside all data, the routine has been ran for the following groups of samples:

- 126 WD (20 pc list).
- 209 WD (25 pc list).

- DA white dwarfs which contains 141 candidates (25 pc list).

- Non- DA white dwarfs contains 68 candidates (25 pc list).

- Hot white dwarfs $\left(T_{eff} \geq 12000\ K°\right)$ with 32 candidates (25 pc list).

- Cool white dwarfs $\left(T_{eff} < 12000\ K°\right)$ with 177 candidates (25 pc list).

The results are listed in Tables (2) to (5), where row 1 represent the mean space velocities, row 2 is the dispersion in velocities, row 3 the eigenvalues, rows 4, 5 and 6 represent the $l,\ m$ and $n$ parameters respectively.

In Table 6, we compare our results as well as the results from different authors as well as inter comparison between our results. We tabulate $\sigma_1,\ \sigma_2,\ \sigma_3,\ (\sigma_2/\sigma_1)$ and the solar $S_\odot$ velocity for our calculations and calculations performed by different authors.

First we focused on the self-comparison between the two sets, 20 pc (Table(2)) and 25 pc (Table (3)) sets. The velocity dispersion $(\sigma_1,\ \sigma_2,\ \sigma_3)$ are comparable for the two samples while the solar velocity are too different. As the 25 pc list exceeds by 167 % higher than the 20 pc list, the solar velocity is lowered by a factor of 22%.

Another comparison between the present results is that between two subsamples represented two different spectral types of white dwarfs, namely DA (141 candidates) and non-DA (68 candidates), Table (4). Here the differences between the two results are significant for both velocity ellipsoid and solar velocity. The difference in the ratio of the velocity dispersion $(\sigma_2/\sigma_1)$ reflect the difference in the initial formation conditions for DA (with rich hydrogen atmosphere and metal core) and non-DA (atmosphere with different chemical compositions) white dwarfs.

Finally, we compared results computed with hot white dwarfs (32 candidates) and cool white dwarfs (177 candidates), Table (5). Also, the results for both the velocity ellipsoid and solar velocity are too different. Here one can guess again the influence of the number of stars on the results, so we cannot draw a conclusion about the variation of the velocity dispersions with the effective temperatures.

Now we turn to the comparisons between our results and that from Wehlau (1957) for dwarfs within 25 pc of the sun. As we see from Table (6) both velocity dispersions and solar velocity are spread over a large range. This could be interpreted by the different method of calculations and number of stars in sample used.

One of the most important quantities in kinematics are Oort's constants. Relation between these constants and the ratio $(\sigma_2/\sigma_1)$ is given by $\sigma_2/\sigma_1 = -B/A-B$. In Table (7) we list values of the constants *A* and *B* adopted from Olling and Merrifield (1998). Column 1 deals with the Oort's constant *A*, column 2 is Oort's constant *B* and column 3 is the ratio $(\sigma_2/\sigma_1)$ calculated with A and B. As we see from the table, the ratio $(\sigma_2/\sigma_1)$ is in the range of 0.65-0.74 which reflects good agreement with our calculations.

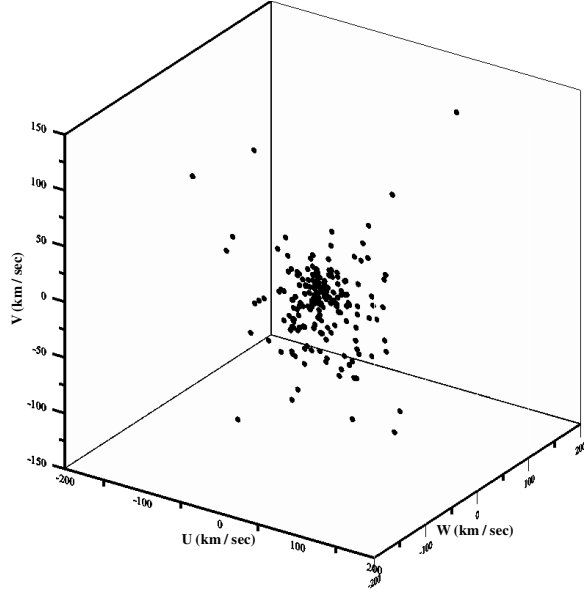

**Fig. 1:** Space velocity distribution of 209 WD (25 pc)

**Table 2:** VEPs for all 126 WD (20 pc), Sion et al. (2009)

| VEPs | | | |
|---|---|---|---|
| $(\overline{U},\overline{V},\overline{W})$ | 1.1373 | -21.9254 | -5.62143 |
| $(\sigma_1,\sigma_2,\sigma_3)$ | 40.9788 | 27.1258 | 34.8816 |
| $(\lambda_1,\lambda_2,\lambda_3)$ | 1679.26 | 735.808 | 1216.73 |
| $(l_1,l_2,l_3)$ deg | 0.319174 | 0.12001 | -0.940067 |
| $(m_1,m_2,m_3)$ | -0.916102 | 0.293066 | -0.273624 |
| $(n_1,n_2,n_3)$ | -0.242664 | -0.94853 | -0.203481 |

**Table 3:** VEPs for all 209 WD (25 pc), Sion et al. (2014)

| VEPs | | | |
|---|---|---|---|
| $(\overline{U},\overline{V},\overline{W})$ | 2.48182 | -18.6593 | -6.94737 |
| $(\sigma_1,\sigma_2,\sigma_3)$ | 40.3025 | 29.6298 | 34.5985 |
| $(\lambda_1,\lambda_2,\lambda_3)$ | 1624.29 | 877.924 | 1197.06 |
| $(l_1,l_2,l_3)$ deg | 0.281297 | 0.381985 | -0.880318 |
| $(m_1,m_2,m_3)$ | -0.892172 | 0.441952 | -0.093314 |
| $(n_1,n_2,n_3)$ | -0.353414 | -0.811644 | -0.465116 |

**Table 4:** VEPs for DA WD and non-DA WD, Sion et al. (2014)

| VEPs | DA (141) | | | non-DA (68) | | |
|---|---|---|---|---|---|---|
| $(\overline{U},\overline{V},\overline{W})$ km/sec | 4.15816 | -17.2078 | -6.28582 | -0.994118 | -21.6691 | -8.31912 |
| $(\sigma_1,\sigma_2,\sigma_3)$ km/sec | 38.5354 | 26.6359 | 31.9558 | 44.0556 | 33.8479 | 40.1042 |
| $(\lambda_1,\lambda_2,\lambda_3)$ km/sec | 1484.98 | 709.473 | 1021.17 | 1940.89 | 1145.68 | 1608.35 |
| $(l_1,l_2,l_3)$ deg | 0.329185 | 0.524158 | -0.785427 | 0.0805911 | 0.178776 | -0.980584 |
| $(m_1,m_2,m_3)$ deg | -0.85574 | 0.51723 | -0.0134785 | -0.959336 | 0.280912 | -0.0276303 |
| $(n_1,n_2,n_3)$ deg | -0.399182 | -0.676559 | -0.618807 | -0.270518 | -0.942936 | -0.194145 |

**Table 5:** VEPs for hot WD and cold WD, Sion et al. (2014)

| VEPs | Hot WD (32) | | | Cool WD (177) | | |
|---|---|---|---|---|---|---|
| $(\overline{U}, \overline{V}, \overline{W})$ km/sec | -0.45312 | -7.16562 | -8.7125 | 3.01243 | -20.7373 | -6.62825 |
| $(\sigma_1, \sigma_2, \sigma_3)$ km/sec | 29.1026 | 17.5027 | 23.3191 | 42.4863 | 30.9974 | 35.9743 |
| $(\lambda_1, \lambda_2, \lambda_3)$ km/sec | 846.96 | 306.343 | 543.781 | 1805.09 | 960.836 | 1294.15 |
| $(l_1, l_2, l_3)$ deg | 0.199976 | 0.0888788 | -0.975761 | 0.310712 | 0.368454 | -0.876185 |
| $(m_1, m_2, m_3)$ deg | 0.532547 | 0.82607 | 0.184386 | -0.902551 | 0.403464 | -0.150397 |
| $(n_1, n_2, n_3)$ deg | 0.822437 | -0.556512 | 0.117862 | -0.298094 | -0.837531 | -0.457909 |

**Table 6:** Velocity dispersions for different spectral types

| Spectral Types | $\sigma_1$ | $\sigma_2$ | $\sigma_3$ | $S_\odot$ | $(\sigma_2/\sigma_1)$ | Reference |
|---|---|---|---|---|---|---|
| 126 WD (2009) | 40.97 | 27.12 | 34.88 | 22.66 | 0.66 | This work |
| 209 WD (2014) | 40.30 | 29.63 | 34.60 | 20.06 | 0.74 | This work |
| 32 hot WD (2014) | 29.10 | 17.50 | 23.31 | 11.28 | 0.60 | This work |
| 177 cold WD (2014) | 42.48 | 30.99 | 35.97 | 21.97 | 0.72 | This work |
| 141 DA WD (2014) | 38.53 | 26.63 | 31.95 | 18.78 | 0.69 | This work |
| 68 non-DA WD (2014) | 44.05 | 33.84 | 40.10 | 23.23 | 0.76 | This work |
| A0-F3 | 20.3 | 9.4 | 9.2 | 13.7 | 0.46 | Wehlau (1957) |
| F4-F8 | 26.5 | 17.3 | 17 | 17.1 | 0.65 | Wehlau (1957) |
| F9-G1 | 25.8 | 18.4 | 20 | 26.4 | 0.71 | Wehlau (1957) |
| G2-G7 | 32.4 | 16.6 | 14.7 | 23.9 | 0.51 | Wehlau (1957) |
| G8-K2 | 28.2 | 15.6 | 11 | 19.8 | 0.55 | Wehlau (1957) |
| K3-K6 | 34.6 | 19.7 | 15.9 | 25 | 0.56 | Wehlau (1957) |
| K8-M2 | 32.1 | 21 | 18.8 | 17.3 | 0.65 | Wehlau (1957) |
| M3-M6 | 31.2 | 23.1 | 16.2 | 23.3 | 0.74 | Wehlau (1957) |

Table 7: Oort's constants.

| A ($km\ s^{-1}\ kpc^{-1}$) | B ($km\ s^{-1}\ kpc^{-1}$) | $\sigma_2/\sigma_1$ |
|---|---|---|
| 14.5 | -12 | 0.65 |
| 12.6 | -13.2 | 0.71 |
| 14.8 | -12.4 | 0.67 |
| 11.3 | -13.9 | 0.74 |

## 4. Conclusion

In summarizing the present paper, velocity dispersions and solar velocity of the white dwarfs within 20 pc and 25 pc are calculated. We have also performed calculations on four subsamples; DA, non-DA, hot and cool white dwarfs. The conclusion reached could be drawn through the following Pointes:

- Increasing the number of white dwarfs of factor$\simeq 2$, the change of the derived parameters are lowered by about 22%.
- Dependence of the derived values on the spectral type of the white dwarfs (DA and non-DA) is clear and reflects the dependence on the chemical composition and consequently on the age of the star.
- We couldn't guess the effect of the effective temperature on the velocity dispersions and solar velocity because of the large difference in the number of the two subsamples (hot and cool white dwarfs).
- Comparison with published parameters for dwarfs within 25 pc of the sun shows a great discrepancies which could be attributed to the type of stars used as well as the method of calculations.